# Photonic toroidal vortex

Chenhao Wan[1,2], Qian Cao[1], Jian Chen[1], Andy Chong[3*], Qiwen Zhan[1*]

[1]School of Optical-Electrical and Computer Engineering, University of Shanghai for Science and Technology, Shanghai 200093, China

[2]School of Optical and Electronic Information, Huazhong University of Science and Technology, Wuhan, Hubei 430074, China

[3]Department of Physics, University of Dayton, 300 College Park, Dayton, Ohio 45469, USA

*Correspondence to: achong1@udayton.edu, qwzhan@usst.edu.cn

**Toroidal vortices are whirling disturbances rotating about a ring-shaped core while advancing in the direction normal to the ring orifice. Toroidal vortices are commonly found in nature and being studied in a wide range of disciplines. Here we report the experimental observation of photonic toroidal vortex as a new solution to Maxwell's equations with the use of conformal mapping. The helical phase twists around a closed loop leading to an azimuthal local orbital angular momentum density. The preparation of such intriguing light field may offer insights of extending toroidal vortex to other disciplines and find important applications in light-matter interaction, optical manipulation, photonic symmetry and topology, and quantum information.**

Toroidal vortices, also known as vortex rings, are intriguing marvels of propagating ring-shaped structures with whirling disturbances rotating about the ring. Toroidal vortices are not uncommon in nature and science. Aquarium visitors are often amazed by dolphins at how they master creating and playing with bubble rings that are air-filled vortex rings propagating in water. The scientific investigation on vortex rings dates back to 1867 when Lord Kelvin proposed the vortex atom model [1]. After one and a half centuries, vortex rings are still being actively investigated in a variety of disciplines. In meteorology, vortex rings of wind, rain and hail are tightly connected to the development of microbursts that pose a great threat to aviation safety [2]. In cardiology, asymmetric redirection of blood flow through the heart resembles a toroidal vortex [3]. In magnetics, experimental observation of vortex rings in a bulk magnet are accomplished just this year, opening up possibilities for studying complex three-dimensional solitons in bulk magnets [4]. In photonics and light science, there have been studies on toroidal dipole and multipole excitation in metamaterials [5]. However, the theory and experimental demonstration on propagating photonic toroidal vortex remains elusive. In this report, we fill this gap based on conformal mapping. Conformal mapping is an angle-preserving transformation that has been utilized to bend optical rays with metamaterials in mapped directions to circumvent hidden objects for achieving optical cloaking [6,7]. In this work, we exploit conformal mapping to reshape a photonic vortex tube into a toroidal vortex. The photonic toroidal vortex is an approximate solution to Maxwell's equations and can propagate without distortion in a uniform medium with anomalous group velocity dispersion. The observation of photonic toroidal vortex, like its counterparts in other disciplines, will spur a wealth of physical mechanisms to explore, such as toroidal electrodynamics, toroidal plasma physics, complex symmetry and topology for light confinement, sensing and manipulation, and interaction of light with metamaterials. The discovery of photonic toroidal vortex may further open up possibilities for the development of novel laser designs, and energy and information transfer methods.

The electric field of an optical wave packet is expressed by the product of the carrier at central frequency and the envelope function. Under scalar, paraxial, and narrow bandwidth approximation, the envelope function that describes the *dimensionless* light field in a uniform medium with anomalous group velocity dispersion is given by [8]

$$\frac{\partial^2 \Psi}{\partial x^2} + \frac{\partial^2 \Psi}{\partial y^2} + \frac{\partial^2 \Psi}{\partial \tau^2} + 2i\frac{\partial \Psi}{\partial z} = 0, \qquad (1)$$

where $\Psi$ is the scalar wave function, $x, y$ are the normalized transverse coordinates, $\tau$ is the normalized retarded time, and $z$ is the propagation distance. As detailed in the supplementary information, a spatiotemporal Laguerre-Gauss tube with vortex line directed in the $y$ direction is a solution to Eq. (1). The iso-intensity surface of a spatiotemporal Laguerre-Gauss tube is labelled *0* as shown in Fig. (1). According to conformal mapping theory, a complex exponential function corresponds to log-polar to Cartesian coordinate transformation. This conformal mapping transforms a line to a circle in two dimensions and performs a tube-to-toroidal mapping in three-dimensional space. Here we adopt an afocal system that consists of two computer generated phase masks to perform optical conformal mapping $(x,y) \mapsto (u,v)$, where $(x,y)$ and $(u,v)$ are the Cartesian coordinates in the planes where the two phase elements are located at. The log-polar to Cartesian coordinate transformation requires $u = b\exp\left(-\frac{x}{a}\right)\cos\left(\frac{y}{a}\right)$ and $v = b\exp\left(-\frac{x}{a}\right)\sin\left(\frac{y}{a}\right)$. The first phase mask performs the mapping from a tube to a toroidal and its phase profile can be derived through stationary phase approximation [9,10]

$$\phi_1(x,y) = \frac{k}{d}\left[-ab\exp\left(-\frac{x}{a}\right)\cos\left(\frac{y}{a}\right) - \frac{x^2+y^2}{2}\right], \qquad (2)$$

where $k$ is the wave number, $d$ is the distance between the two phase masks, $a$ and $b$ are used to adjust the beam size and position in the $(u,v)$ plane. The simulated evolution process of the mapping from a tube to a vortex ring in free-space propagation as labelled from *1* to *5* is shown in Fig. 1. The second phase mask recollimates the light. In time-reversal direction, the second phase mask performs a Cartesian to log-polar coordinate transformation. Similarly, its phase profile can be derived as

$$\phi_2(u,v) = \frac{k}{d}\left[-au\ln\frac{\sqrt{u^2+v^2}}{b} + av\arctan\left(\frac{v}{u}\right) + au - \frac{u^2+v^2}{2}\right]. \quad (3)$$

After travelling through the two phase masks, the spatiotemporal vortex tube will be mapped into a photonic toroidal vortex.

The toroidal vortex as a solution to Eq. (1) in the remapped coordinates can be approximately expressed as

$$\Psi = \left(\frac{\sqrt{(r_\perp - r_0)^2 + \tau^2}}{\sqrt{z_0}}\right)^l \exp\left(-\frac{(r_\perp - r_0)^2 + \tau^2}{2z_0}\right)\exp\left(-il\tan^{-1}\left(\frac{\tau}{r_\perp - r_0}\right)\right), \quad (4)$$

where $r_\perp = \sqrt{u^2 + v^2}$, $r_0$ is the radius of the circular vortex line, $l$ is an integer, and $z_0$ is a constant. Figure 2 shows the iso-intensity surface of a photonic toroidal vortex in different perspective views. The photonic toroidal vortex possesses a three-dimensional phase structure that rotates around a closed loop forming a ring-shaped vortex line while advancing in the direction perpendicular to the ring orifice at the speed of light. The rotating spatiotemporal spiral phase is represented by colorful painting and the direction of local orbital angular momentum (OAM) density is marked with arrows.

The experimental apparatus begins with a mode-locked fiber laser that emits a chirped pulse (Fig. 3). The light from the source splits into a reference pulse and a signal pulse. The reference pulse is dechirped to transform-limited pulse through a grating pair. A reflective mirror is mounted on a precision stage after the grating pair so that the path length difference between the reference pulse and the signal pulse can be accurately controlled. The signal pulse propagates through a two-dimensional pulse shaper that consists of a grating, a cylindrical lens and a reflective programmable spatial light modulator (SLM 1). SLM 1 situates in the spatial frequency - temporal frequency plane and applies a controlled spiral phase to the signal pulse. After a spatiotemporal Fourier transform, the signal pulse is converted to a wave packet carrying a spatiotemporal vortex[11-14]. The spatiotemporal vortex pulse then travels through an afocal cylindrical beam expander and stretches in the direction of the vortex line. SLM 2 and SLM 3 compose an afocal mapping system. They are programmed with the phase profiles as given by Eq. (2) and Eq. (3), respectively. The parameters *a* and *b* are properly chosen so that the light occupies a great portion of the liquid crystal area. The stretched spatiotemporal vortex pulse transforms into a toroidal vortex pulse through the conformal mapping system.

To characterize the three-dimensional field information of the toroidal vortex pulse, the transform-limited reference pulse is combined with the toroidal vortex pulse by a beam splitter (BS 4) and the interference patterns are recorded by a charged-coupled device (CCD) camera. The reference pulse is considerably shorter (~90 fs) than the toroidal vortex pulse (~3 ps), therefore, by the use of a precision stage and a movable mirror, the reference pulse interferes with each temporal slice of the toroidal vortex pulse. Through utilizing the cross-correlation method [15,16], the amplitude and phase information of each slice can be retrieved from the interference patterns. Consequently, the three-dimensional iso-intensity profile of the toroidal vortex pulse can be reconstructed as shown in Fig. 4. Semitransparency is applied so that the ring orifice and the hidden ring-shaped vortex core are clearly shown. The vortex core is colored in red and the local OAM density is denoted by green arrows. In order to present the spiral phase rotating about the vortex core, three slices are taken in the radial direction of the ring as marked numerically. The phase profiles of the three slices in local coordinates are reconstructed and shown in Fig. 4c. All three images show a spiral phase of topological charge 1. The intensity and phase information demonstrate that the generated pulse is indeed a toroidal vortex.

In conclusion, we experimentally demonstrate the generation of photonic toroidal vortex as a new solution to Maxwell's equations through optical conformal mapping. The preparation of photonic vortex rings is of great importance for studying toroidal electrodynamics, light-matter interaction, and photonic symmetry and topology. The concept and conformal mapping method can be readily extended to other spectrums such as electron beams, X-rays, and even mechanical waves such as acoustics, hydrodynamics and aerodynamics, opening up new opportunities for studying vortex rings in a wide range of disciplines.

**Funding**

We acknowledge the support from the National Natural Science Foundation of China (NSFC) (92050202, 61875245, 61805142), Shanghai Science and Technology Committee (19060502500), and Wuhan Science and Technology Bureau (2020010601012169).

**Author Contributions**

C.W., A.C. and Q.Z. proposed the original idea and performed the theoretical analysis. C.W. performed all experiments and data analysis. Q.C. and J.C. contributed in developing the measurement method. Q.Z. guided the theoretical analysis and supervised the project. All authors contributed to writing the manuscript.

**Competing interests**

Authors declare no competing interests.

**Data and materials availability**

All data presented in the report are available upon reasonable request.

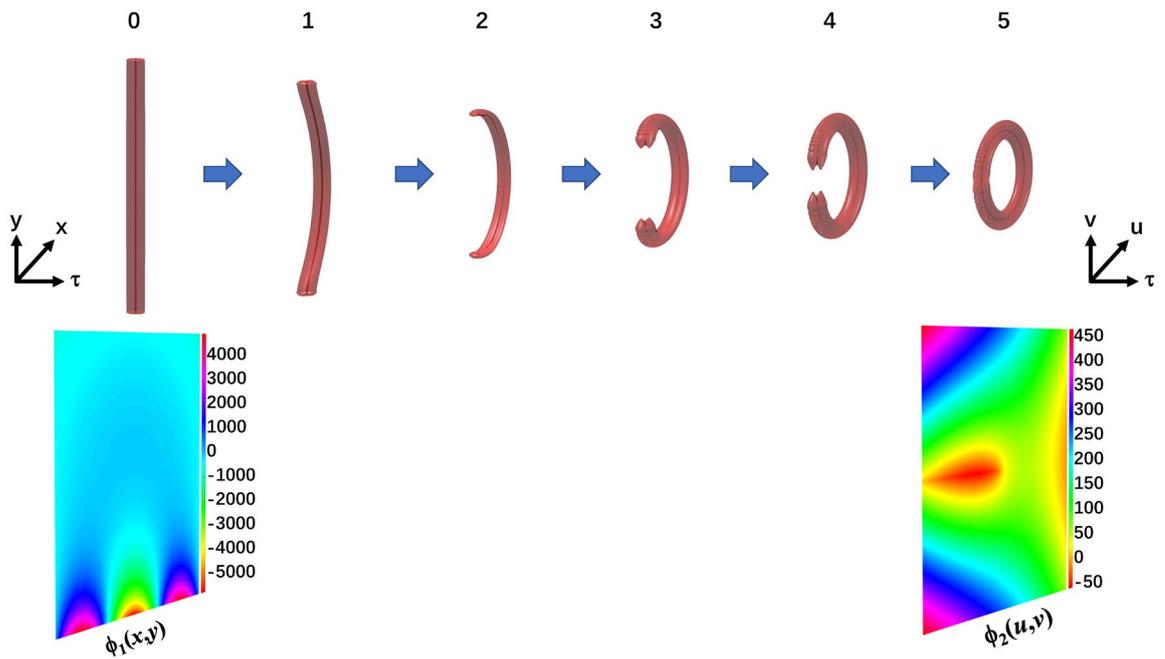

**Fig. 1. Simulation of the conformal mapping from a spatiotemporal vortex tube to a vortex ring.** The spatiotemporal vortex tube acquires phase ϕ$_1$(x,y) and begins the transformation process in free-space repropagation. After the completion of vortex ring formation, a second phase mask ϕ$_2$(u,v) is applied to collimate the light.

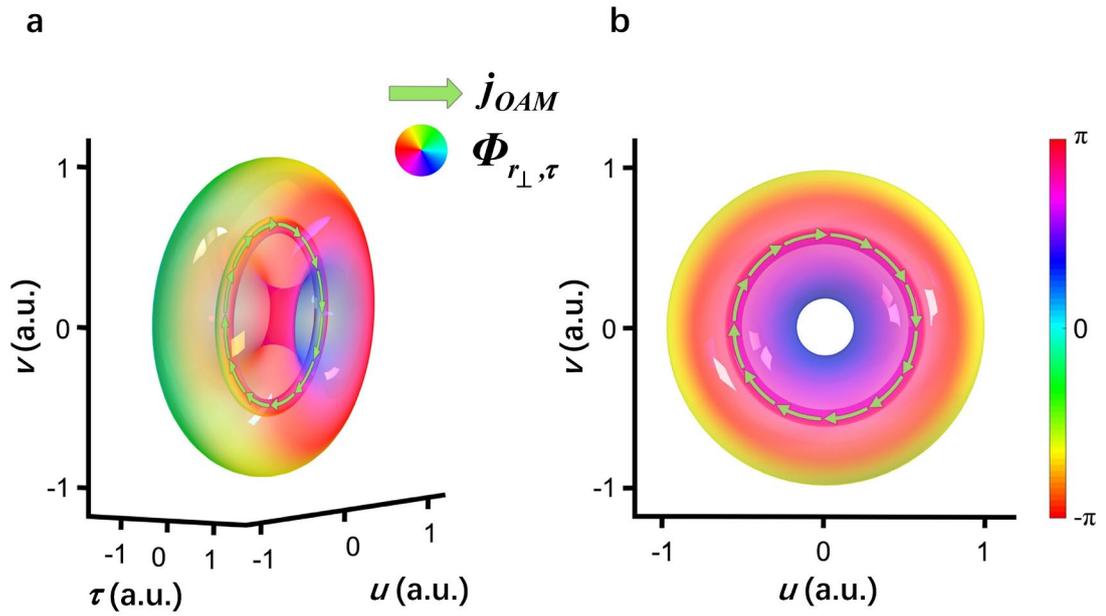

**Fig. 2. Intensity and phase information of a simulated photonic toroidal vortex. a,** Side view of three-dimensional iso-intensity profile of a photonic toroidal vortex. The colorful painting denotes the rotating spatiotemporal spiral phase ($\Phi_{r_\perp,t}$). The topological charge of the spiral phase is 1. The direction of local OAM density ($j_{OAM}$) is marked with arrows. **b,** Front view of the iso-intensity profile of a photonic toroidal vortex.

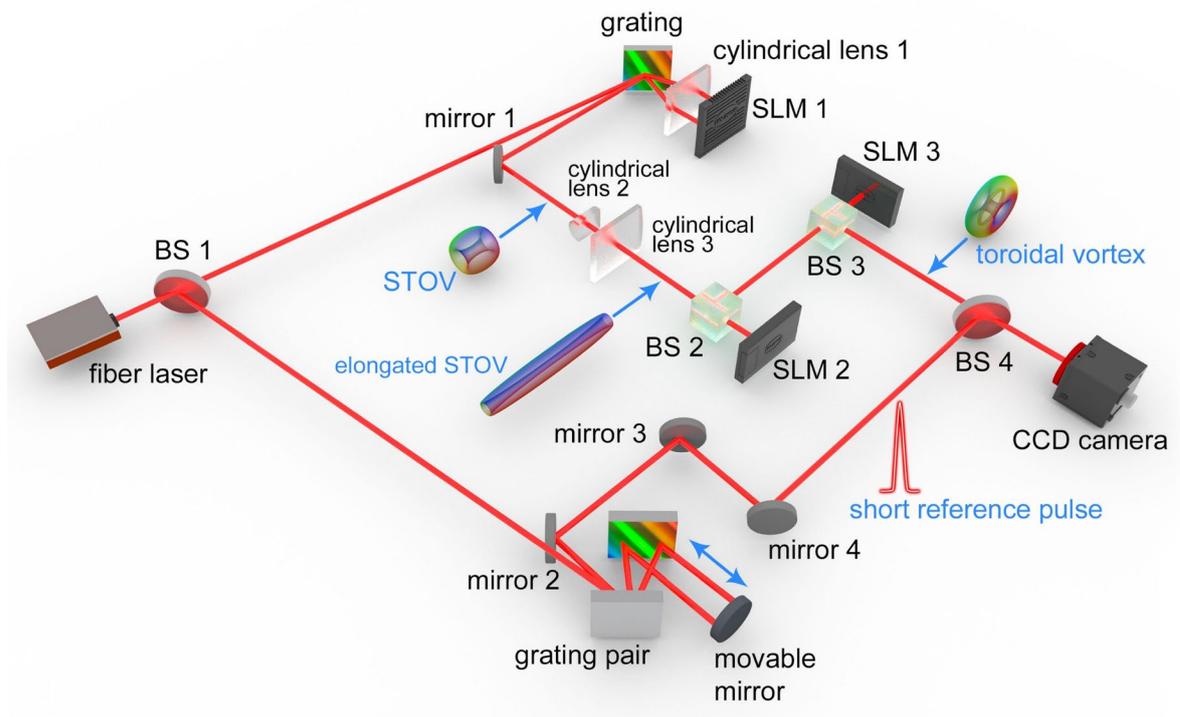

**Fig. 3. Schematic of the experimental apparatus.** A chirped pulse from the laser source splits to a signal pulse and a reference pulse. The signal pulse transforms to a spatiotemporal vortex (STOV) pulse after a two-dimensional pulse shaper. The spatiotemporal vortex is stretched along the vortex line and then converted to a toroidal vortex through an afocal conformal mapping system. The toroidal vortex is characterized by interference with the dechirped reference pulse.

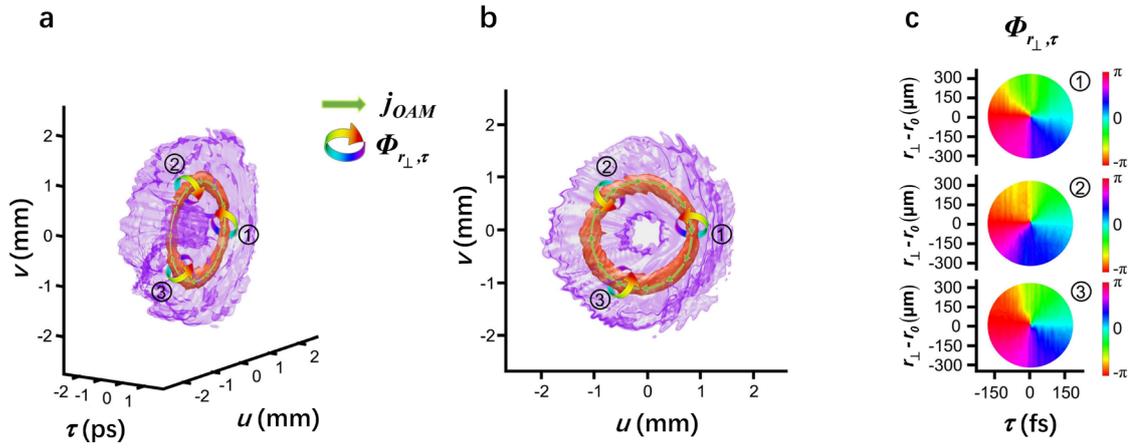

**Fig. 4. Intensity and phase information of an experimentally generated photonic toroidal vortex. a-b,** Three-dimensional iso-intensity profile of the photonic toroidal vortex in different views. The spatiotemporal spiral phase is ($\Phi_{r_\perp,\tau}$) marked with curved colorful arrows and the local OAM density ($j_{OAM}$) is marked with light green arrows. **c,** Circulating spiral phase in the radial-temporal plane. The topological charge of the spiral phase is 1.